\documentclass[amsmath,amssymb,showpacs,twocolumn,superscriptaddress,pr]{revtex4-1}

\usepackage{graphicx,bm,color,subfigure}
\usepackage[latin9]{inputenc}
\usepackage{colortbl}
\usepackage {color}
\usepackage{xcolor}
\usepackage{bm}
\usepackage{amsmath}
\usepackage{hyperref}
\hypersetup{colorlinks,
	citecolor=blue}
\usepackage{txfonts}
\begin{document}
	
	\title{Hydrogen Doping Induced $p_x\pm ip_y$ Triplet Superconductivity in Quasi-One-Dimensional K$_2$Cr$_3$As$_3$}
	
	\author{Ming Zhang}
	\email{mingz@zstu.edu.cn}
	\thanks{These two authors contributed equally to this work.}
	\affiliation{Department of Physics, Zhejiang Sci-Tech University, Hangzhou, Zhejiang, 310018 P. R. China}
	
	\author{Chen Lu}
	\thanks{These two authors contributed equally to this work.}
	\affiliation{New Cornerstone Science Laboratory, Department of Physics, School of Science, Westlake University, Hangzhou 310024, Zhejiang, China}
	
	\author{Yajiang Chen}
	\affiliation{Department of Physics, Zhejiang Sci-Tech University, Hangzhou, Zhejiang, 310018 P. R. China}
	
	\author{Yunbo Zhang}
	\affiliation{Department of Physics, Zhejiang Sci-Tech University, Hangzhou, Zhejiang, 310018 P. R. China}
	
	\author{Fan Yang}
	\email{yangfan\_blg@bit.edu.cn}
	\affiliation{School of Physics, Beijing Institute of Technology, Beijing 100081, China}

	\begin{abstract}
		Quasi-one-dimensional (Q1D) Cr-based pnictide K$_2$Cr$_3$As$_3$ has aroused great research interest due to its possible triplet superconducting pairing symmetry. Recent experiments have shown that incorporating hydrogen atoms into K$_2$Cr$_3$As$_3$ would significantly change its electronic and magnetic properties. Hence, it's necessary to investigate the impact of hydrogen doping in superconducting pairing symmetry of this material. Employing the hydrogen as a non-trivial electron-doping, our calculations show that, different from the $p_z$-wave obtained without hydrogen, the system exhibits $p_x\pm ip_y$ pairing superconductivity under specific hydrogen doping. Specifically, we adopt the random-phase-approximation approach based on a six-band tight-binding model equipped with multi-orbital Hubbard interactions to study the hydrogen-doping dependence of the pairing symmetry and superconducting $T_c$. Under the rigid-band approximation, our pairing phase diagram shows that the spin-triplet pairing state is dominated in the hydrogen-doping regime $x\in (0,0.7)$. Particularly, the $T_c\sim x$ curve shows a peak at the 3D-quasi-1D Lifshitz transition point, and the pairing symmetry around this doping level is $p_x\pm ip_y$. The physical origin of this pairing symmetry is that the density of states is mainly concentrated at $k_x(k_y)$ with large momentum. Due to the three-dimensional character of the real material, this $p_x\pm ip_y$-wave superconducting state possesses point gap nodes. We further provide experiment prediction to identify this triplet $p_x\pm ip_y$-wave superconductivity.

	\end{abstract}

	

	\maketitle
	

	\section{Introduction}
	
	Since 2015, the discovery of the superconductivity in quasi-one-dimensional (Q1D) Cr-based compounds A$_2$Cr$_3$As$_3$ (A = K, Rb, Cs) has generated immediate and continued research interest \cite{Bao:15,Tang:15a,Tang:15,Pang:15,Zhi:15,Adroja:15,Kong:15,Yang:15,Balakirev:15,Wang:15,Pang:16,Cao:17,Adroja:17,Taddei:17,Zhao:18,Mu:18a,Mu:18,Luo:19}. These compounds consist of alkali-metal-atom-separated [(Cr$_3$As$_3)^{2-}$]$_\infty$ double-walled subnanotubes with the low-energy degrees of freedom dominated by the Cr $3d$ orbitals \cite{Jiang:15,Hu:15}, which are proposed to be strongly correlated \cite{Zhi:15,Yang:15,Taddei:17,Wu:15,Zhang:16,Zhou:17,Miao:16,Dai:15,Wu:1507}, implying an electron-interaction-driven pairing mechanism. Evidence for unconventional SC in K$_2$Cr$_3$As$_3$ and its analogues \cite{Tang:15a,Tang:15,Pang:15} has been accumulated from various experiments \cite{Zhi:15,Yang:15,Adroja:15,Pang:15,liu2016,Zuo2017}. In particular, a novel spin-triplet pairing was theoretically proposed \cite{Wu:15,Jiang:15,Zhang:16,Zhou:17,Alaska}, in accordance with the ferromagnetic spin fluctuations \cite{Yang:15, Jiang:15}. Recently, the nuclear magnetic resonance (NMR) measurements of the Knight shift show spin triplet superconductivity behavior in K$_2$Cr$_3$As$_3$ \cite{Jie2021}. Besides, spin relaxation rate study has revealed point nodes in the gap function \cite{Yang:15,Luo:19}. Therefore, superconducting pairing symmetry in this system could be $p_x\pm ip_y$-wave. Such a state breaks time reversal symmetry and is consistent with zero-field muon spin resonance ($\mu$SR) measurement that revealed evidence for a spontaneous appearance of a weak internal magnetic field below $T_c$ \cite{Adroja:15}.

	Slightly after the synthesization of the A$_2$Cr$_3$As$_3$ (233) family, A$_1$Cr$_3$As$_3$ (133) was obtained by removing half of the A$^+$ ions and it exhibits superconductivity after doped with a certain percentage of hydrogen. Inspired by this, recent experiments intercalated hydrogen in K$_2$Cr$_3$As$_3$ to obtain K$_2$Cr$_3$As$_3$H$_x$. Measuring its magnetic properties shows the presence of an AFM transition \cite{Li2023}. Note that, without hydrogen doping, the system has a ferromagnetic spin fluctuation \cite{cuono2021,Jie2021}, and some theoretical calculations show that the superconductivity pairing symmetry is triplet $p_z$-wave \cite{Wu:15,Zhang:16,Hu:15}. Thus it is necessary to reconsider the superconducting pairing symmetry and magnetic fluctuations in this hydrogen-doped system.

	The DFT-based calculations \cite{Li2023} show that the chemical reaction between K$_2$Cr$_3$As$_3$ and H$_2$ will form K$_2$Cr$_3$As$_3$H with similar quasi-1D structure as that of K$_2$Cr$_3$As$_3$, but with the hydrogen atoms intercalated at the center of Cr octahedra in the [(Cr$_3$As$_3$)$^{2-}$]$_{\infty}$ subnanotubes. In the aspect of band structure \cite{Li2023}, the role of the intercalated hydrogen atoms mainly lie in the rise of the Fermi energy $E_F$, besides modest distortions to the bands near $E_F$. Therefore, we can say that in K$_2$Cr$_3$As$_3$H$_x$, H has metallic bonding and acts as an electron donor. Furthermore the H concentration $x$ in the material is experimentally tunable \cite{Li2023}. While the DFT results for $x=1$ yield in-plane-antiferromagnetic ordered ground state \cite{Li2023}, those for $x=0$ suggest non-magnetic ground state with short-ranged ferromagnetic \cite{Wu:15} spin fluctuations, which  might mediate spin-triplet superconductor. Therefore, the phase diagram in K$_2$Cr$_3$As$_3$H$_x$ via tuning $x$ is like those of the cuprates and the iron-pnictide superconductors wherein magnetic order states are usually found to be proximate to the SC, suggesting the relevance of the e-e interaction driven pairing mechanism. However, detailed theoretical studies about this phase diagrams are still missing.
	
	In this article, we study the pairing symmetry of the K$_2$Cr$_3$As$_3$H$_x$ via the random-phase-approximation (RPA) approach \cite{RPA1,RPA2,RPA3,Kuroki2008,Scalapino2009,Scalapino2011,Liu2013,Liu2018,ZhangLD:19}, adopting the six-band tight-binding (TB) model proposed in Ref. \cite{Wu:15}. Based on the band structure for $x=0$, we use the rigid-band approximation to study the $x$- dependence of the pairing symmetry and the superconducting $T_c$ in the regime $x\in (0,0.7)$. Our results yield that the hydrogen doping induces a transition in the pairing symmetry of the system, from a $p_z$-wave type to a $p_x\pm ip_y$-wave type. Particularly, when the $p_x\pm ip_y$-wave leads the pairing symmetry, $T_c$ reaches its peak. Investigating the Fermi surface (FS) we found this high $T_c$ is obtained at the 3D-quasi-1D Lifshitz-transition doping level. To understand the physical origin of this superconducting phase transition, we investigate the distribution of the density of states (DOS) over the Fermi surface at different dopings. It shows that for dopings near 0.3, where the $p_x\pm ip_y$-wave leads the pairing symmetry, the DOS reaches its peak at larger $k_x(k_y)$ compared with $k_z$. For other dopings where $p_z$-wave is the leading pairing symmetry, the DOS reaches its peak at larger $k_z$. We further discuss experimental verification of this $p_x\pm ip_y$-wave superconducting state with point gap nodes.

	The rest of this paper is organized as follows: In Sec. \ref{sec:TB}, we construct the effective TB model and introduce the details of the RPA approach engaged in our calculations. In Sec. \ref{sec:FTA}, we analyze the magnetic fluctuations affecting the superconducting pairing through the dependence of the susceptibility on doping. In Sec. \ref{phase_diagram} we calculate the dependence of superconducting pairing symmetry on hydrogen doping using the RPA approach. In Sec. \ref{p+ip}, we further propose the physical origin and experimental behaviors of the obtained $p_x\pm ip_y$-wave superconductivity. A summary is given in Sec. \ref{sec:conclusion} together with some discussions about possible experimental implications.

	\section{MODEL AND APPROACH}
	\label{sec:TB}
	\subsection{The TB band structure}
	\label{subsec:LDA}
	
	\begin{figure}[htbp]
		\centering
		\includegraphics[width=0.5\textwidth]{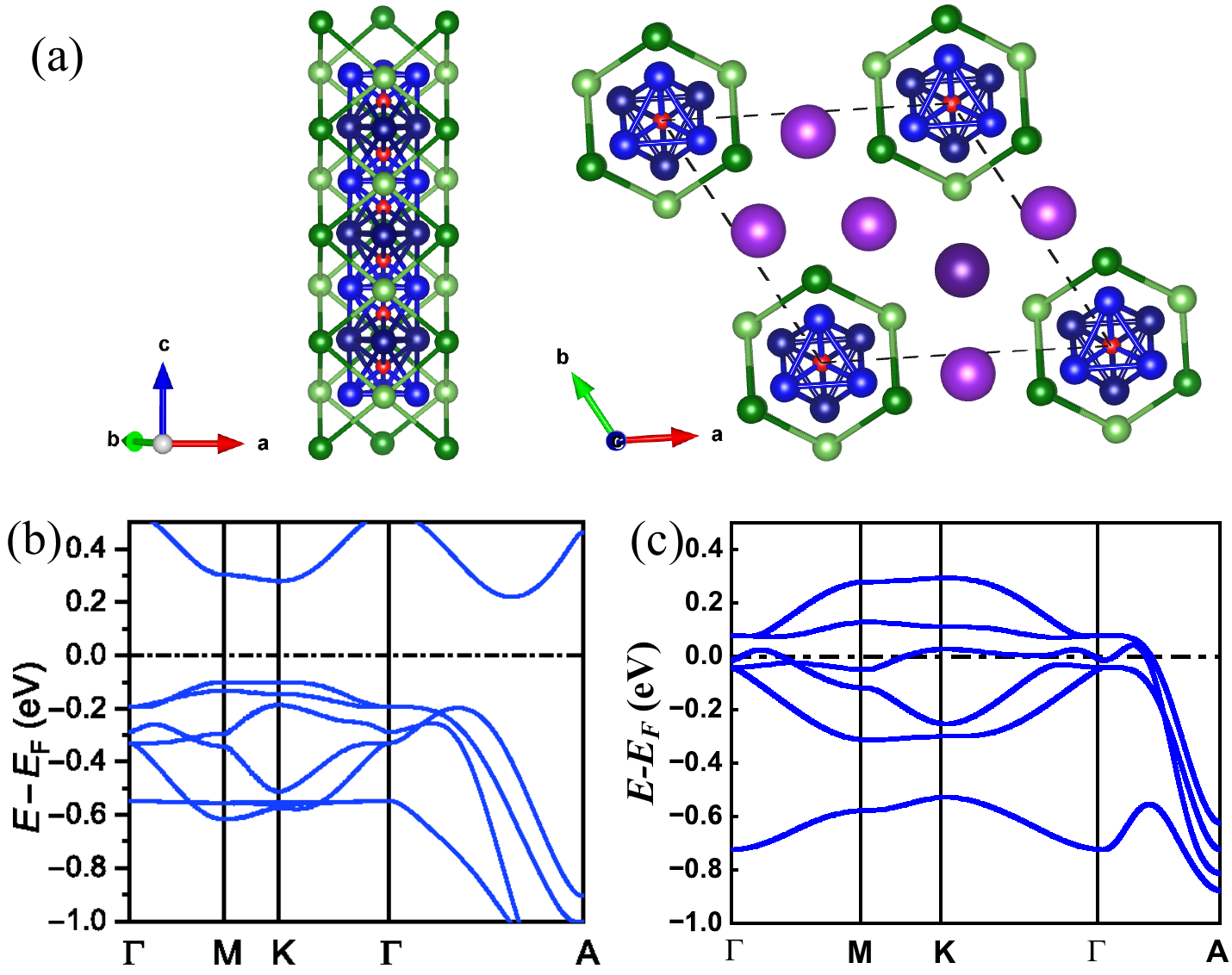}
		\caption{(color online). Structural and band characterization of K$_2$Cr$_3$As$_3$H$_x$. (a) The crystal structure projected along
			b and c directions, respectively. (b) Band structure of K$_2$Cr$_3$As$_3$H along the high-symmetry lines. (c) Band structure of K$_2$Cr$_3$As$_3$ in the same energy range as (b).}
		\label{crystal}
	\end{figure}
	
	K$_2$Cr$_3$As$_3$H$_x$ crystallizes in a hexagonal lattice with $a=10.0242(5)\mathring{\mathrm{A}}$ and $c=4.2501(8)\mathring{\mathrm{A}}$, with are $0.41\%$ and $0.46\%$ larger than the counterparts of K$_2$Cr$_3$As$_3$ \cite{Li2023}. The point group of this lattice is the same with the original lattice and it is $D_{3h}$, which includes a $C_3$ rotation about the $z$ axis and a mirror reflection about the $xy$ plane. The constituent atoms locate in the two crystalline planes with $z=0$ and $z=0.5$, as illustrated in Fig. \ref{crystal}(a). The hydrogen atoms is intercalated into the center of the CrAs tubes between the stacked CrAs layers, creating a chain of hydrogen centered in each tube. Compared with the K$_2$Cr$_3$As$_3$H's low-energy band structure from the DFT calculations shown in Fig. \ref{crystal}(b), K$_2$Cr$_3$As$_3$ shows a similar shape, with only modest distortion near the Fermi level that is relatively decreased by about 0.3 eV (Fig. \ref{crystal}(c)). Therefore, the inserted hydrogen atoms in K$_2$Cr$_3$As$_3$H$_x$ can be well viewed as effective electron donors, consistent with previous results \cite{Li2023}.

	\begin{figure}[htbp]
		\centering
		\includegraphics[width=0.5\textwidth]{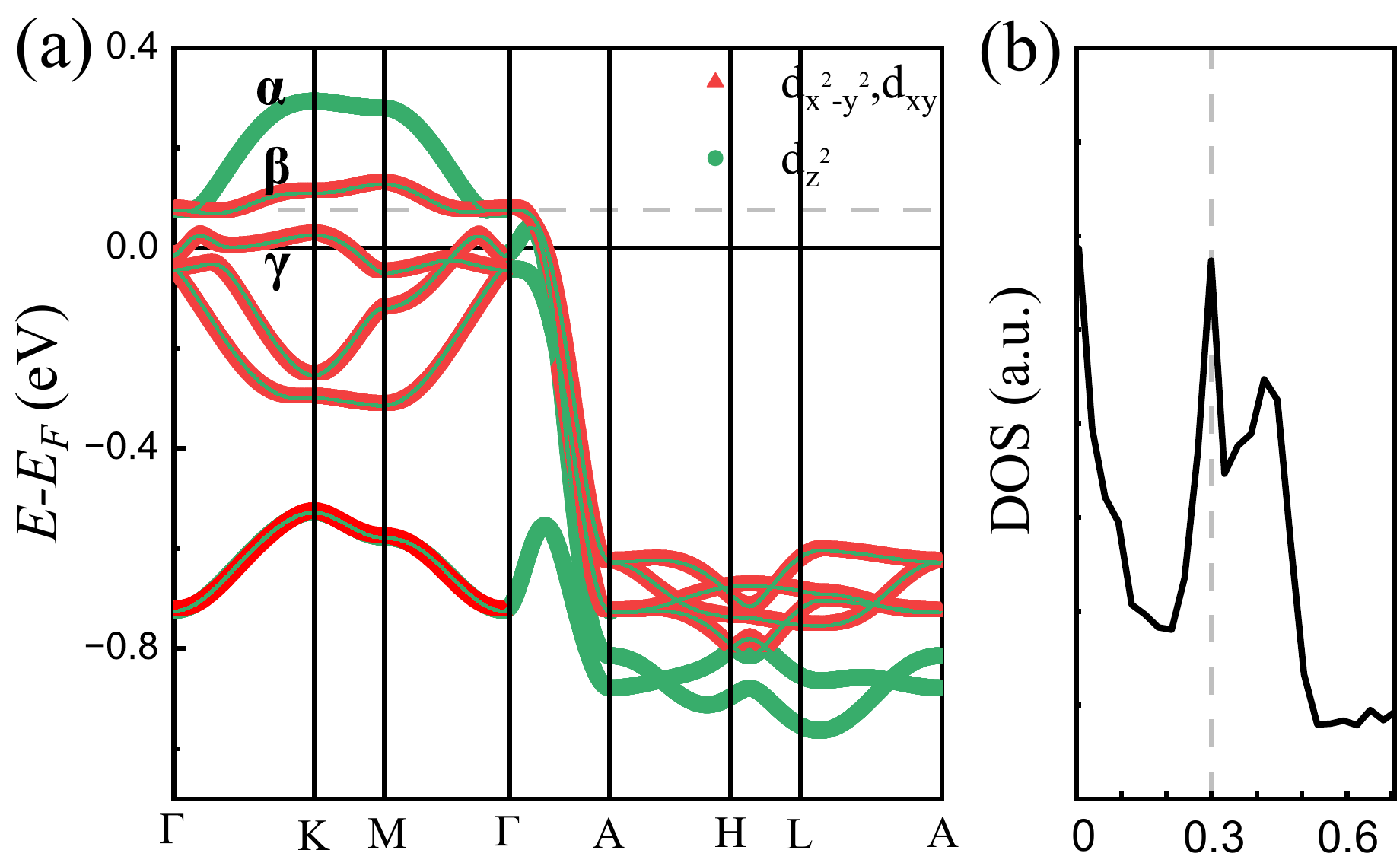}
		\caption{(color online). Band structure (a) and density of energy state (b) of K$_2$Cr$_3$As$_3$H$_x$ from the six-band TB model. Hydrogen doping at the grey dashed line in (a) is 0.3. }
		\label{band}
	\end{figure}

	As the IOP/CLK magnetic order calculated by DFT and a non-negligible out-of-plane coupling was estimated in Heisenberg model \cite{Li2023}, we adopt a six-band TB model with both Cr1 and Cr2 involved which can reflect both the intrachain and interchain magnetic properties of the system. This model was first proposed in Ref. \cite{Wu:15}. Following Ref. \cite{Wu:15}, the obtained TB Hamiltonian in momentum space can be expressed as,
	\begin{align}
		H_{{\rm TB}}
		=\sum_{\bm{k}mn\mu\nu\sigma}h_{\mu\nu}^{mn}(\bm{k})
		c^{\dagger}_{m\mu\sigma}(\bm{k})c_{n\nu\sigma}(\bm{k}),
	\end{align}
	Here $m/n=A,B$ labels sublattice, $\mu,\nu=1,\cdots, 3$ indicating the orbital indices, containing the $d_{z^2}$, $d_{x^2-y^2}$ and $d_{xy}$ orbitals, and $\sigma=\uparrow,\downarrow$ labels spin. $c^{\dagger}_{m\mu\sigma}(\bm{k})(c_{n\nu\sigma}(\bm{k})$ creates(annihilates) a spin-$\sigma$ electron in the orbital $\mu(\nu)$ in the $m$th($n$th) sublattice with momentum $\bm{k}$. The elements of the $h(\bm{k})$ matrix of this six-band tight-binding model are given in Ref. \cite{Wu:15}.

	Although the above-provided band structure and TB model are only accurately applicable to K$_2$Cr$_3$As$_3$, we take  rigid-band approximation and adopt it to describe the band structure of K$_2$Cr$_3$As$_3$H$_x$, with only the chemical potential tuned according to the variation of $x\in (0,0.7)$. Note than,  too high $x$ might invalidate the rigid-band approximation as the band structure we adopt is for $x = 0$. Besides, the insulating state instead of SC is experimentally detected for $x$ close to 1 \cite{Li2023}.
	
\begin{figure}[htbp]
		\centering
		\includegraphics[width=0.45\textwidth]{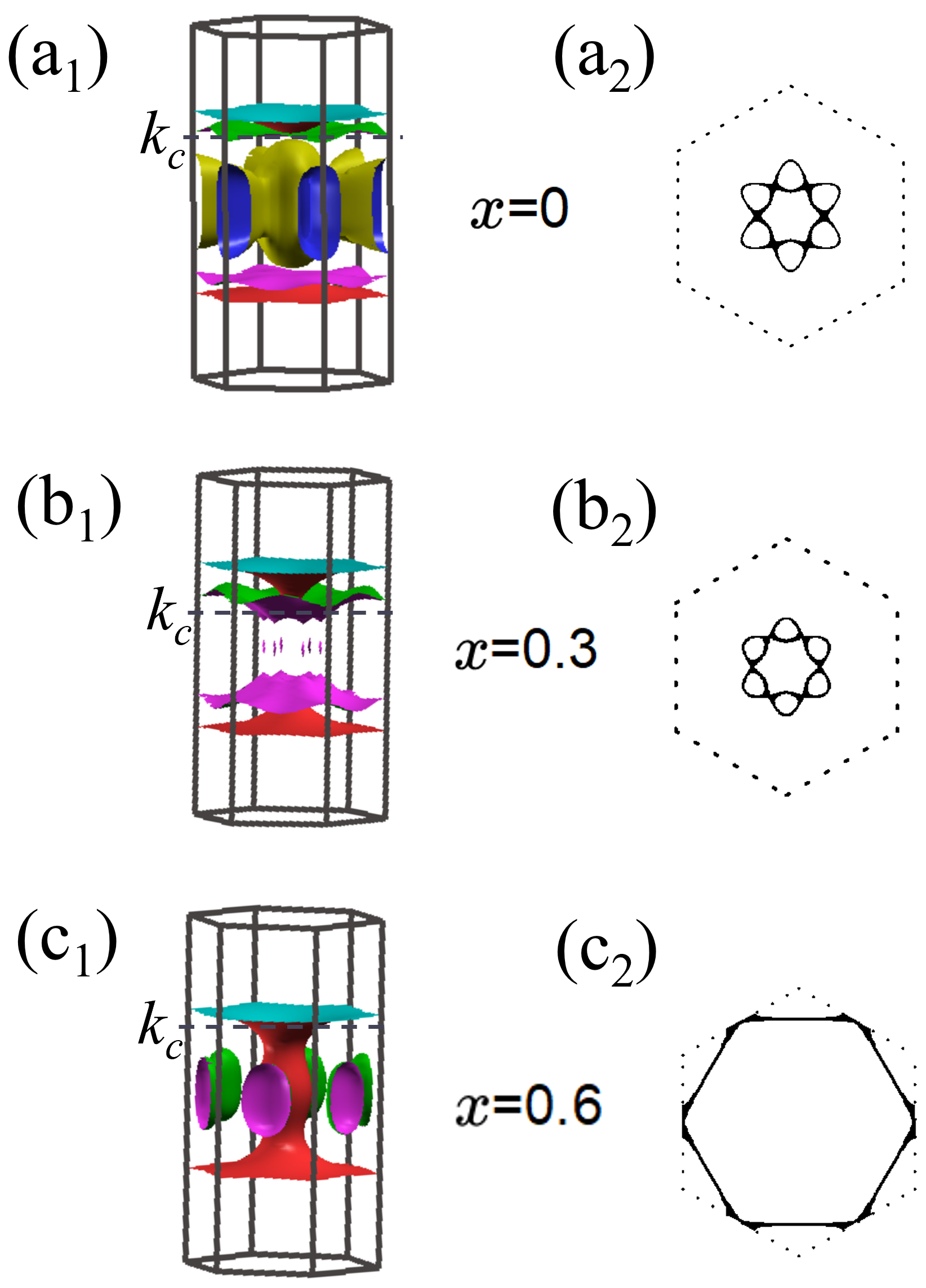}
		\caption{(color online). 3D FSs and 2D cuts of the FSs on $k_z=k_c$ plane for K$_2$Cr$_3$As$_3$H$_x$ with typical doping levels (a) $x=0$, (b) $x=0.3$, and (c) $x=0.6$ under the rigid-band approximation. Here $k_c$ marked in 3D FSs is the corresponding $k_z$ of 2D cuts. }
		\label{fs}
	\end{figure}

	Considering the presence of H deficiencies in experiments and a peak in the DOS at 0.3 doping during the rise in chemical potential as shown in Fig. \ref{band}(b), it is necessary to investigate how the FS are transformed in this process. We examined the doping dependence of FS and show three specific doping levels in Fig \ref{fs}, i.e., $x=0$(Fig. \ref{fs}($a_1$)),$x=0.3$(Fig. \ref{fs}($b_1$)) and $x=0.6$(Fig. \ref{fs}($c_1$)). Although more doping dependence of FS have not been shown due to space limitations, one can find that the topology of the Fermi surface has significantly changed during doping. Particularly, a 3D-quasi-1D Lifshitz-transition occurred when doping around $x=0.3$. Further, we marked the Van Hove Singularities' (VHS')s $k_z$ with $k_c$ of each 3D FS, and  made corresponding 2D cut as shown in the right column of Fig \ref{fs}.
	
	\subsection{The RPA approach}
	\label{sec:RPA}
	
	We first define the following bare susceptibility tensor in the normal state for the non-interacting case:
	\begin{align}\label{chi0}
		\chi^{(0)}_{pqst}(\bm{k},\tau)\equiv
		&\frac{1}{N}\sum_{\bm{k}_1\bm{k}_2}\left\langle
		T_{\tau}c_{p}^{\dagger}(\bm{k}_1,\tau)
		c_{q}(\bm{k}_1+\bm{k},\tau)\right.                      \nonumber\\
		&\left.\times c_{s}^{\dagger}(\bm{k}_2+\bm{k},0)
		c_{t}(\bm{k}_2,0)\right\rangle_0,
	\end{align}
	Here $\langle\cdots\rangle_0$ denotes the thermal average for the noninteracting system, $T_{\tau}$ denotes the imaginary time-ordered product, and the tensor indices $p,q,s,t=1,\cdots,6$ denote the orbital-sublattice indices. Fourier transformed to the imaginary frequency space, the bare susceptibility can be expressed by the following explicit formulism:
	\begin{align}\label{chi0e}
		\chi^{(0)}_{pqst}(\bm{k},i\omega_n)
		=&\frac{1}{N}\sum_{\bm{k}'\alpha\beta}
		\xi^{\alpha}_{t}(\bm{k}')
		\xi^{\alpha*}_{p}(\bm{k}')
		\xi^{\beta}_{q}(\bm{k}'+\bm{k})                         \nonumber\\
		&\times\xi^{\beta*}_{s}(\bm{k}'+\bm{k})
		\frac{n_F(\varepsilon^{\beta}_{\bm{k}'+\bm{k}})
			-n_F(\varepsilon^{\alpha}_{\bm{k}'})}
		{i\omega_n+\varepsilon^{\alpha}_{\bm{k}'}
			-\varepsilon^{\beta}_{\bm{k}'+\bm{k}}}.
	\end{align}
	where $\alpha,\beta=1,\cdots,6$ are band indices, $\varepsilon^{\alpha}_{\bm{k}}$ and $\xi^{\alpha}\left(\bm{k}\right)$ are the $\alpha$-th eigenvalue (relative to the chemical potential $\mu_c$) and eigenvector of the TB model, respectively, and $n_F$ is the Fermi-Dirac distribution function.
	
	To study superconductivity pairing phase diagram, we adopt the following extended Hubbard model Hamiltonian:
	\begin{align}\label{model}
		H=&H_{\text{TB}}+H_{int}\nonumber\\
		H_{int}=&U\sum_{i\mu}n_{i\mu\uparrow}n_{i\mu\downarrow}+
		V\sum_{i,\mu<\nu}n_{i\mu}n_{i\nu}+J_{H}\sum_{i,\mu<\nu}                   \nonumber\\
		&\Big[\sum_{\sigma\sigma^{\prime}}c^{+}_{i\mu\sigma}c^{+}_{i\nu\sigma^{\prime}}
		c_{i\mu\sigma^{\prime}}c_{i\nu\sigma}+(c^{+}_{i\mu\uparrow}c^{+}_{i\mu\downarrow}
		c_{i\nu\downarrow}c_{i\nu\uparrow}+h.c.)\Big]
	\end{align}
	Here, the interaction parameters $U$, $V$, and $J_H$ denote the intra-orbital, inter-orbital Hubbard repulsion, and the Hund's rule coupling (as well as the pair hopping) respectively, which satisfy the relation $U=V+2J_H$.

	When the Hubbard interaction in Eq. (\ref{model}) is included, we can explicitly calculate the spin $(s)$ and charge $(c)$ susceptibilities following the standard multi-orbital RPA approach \cite{RPA1,RPA2,RPA3,Kuroki2008,Scalapino2009,Scalapino2011,Liu2013,Liu2018,ZhangLD:19}. At the RPA level, the renormalized spin and charge susceptibilities of the system read
	\begin{align}\label{chisce}
		\chi^{(s,c)}(\bm{k},i\omega_n)=[I\mp\chi^{(0)}(\bm{k},i\omega_n)
		U^{(s,c)}]^{-1}\chi^{(0)}(\bm{k},i\omega_n),
	\end{align}
	Here the nonzero elements $U^{(s,c)\mu\nu}_{\theta\xi}$ of $U^{(s,c)}$ satisfy $\mu,\nu,\theta,\xi\leq 3$ or $>3$ simultaneously, which are as follow,
	\begin{align}
		U^{(s(c))\mu\nu}_{\theta\xi}=\left\{
		\begin{array}{ll}
			U(U),  & \mu=\nu=\theta=\xi; \\
			J_H(2V-J_H), & \mu=\nu\neq\theta=\xi; \\
			J_H(J_H), & \mu=\theta\neq\nu=\xi; \\
			V(2J_H-V),  & \mu=\xi\neq\theta=\nu.
		\end{array}
		\right.
	\end{align}
	In Eq. (\ref{chisce}), $\chi^{(s,c,0)}(\bm{k},i\omega_n)$ and $U^{(s,c)}$ are operated as $6^{2}\times 6^{2}$ matrices (see for example in Ref. \cite{Liu2013}).

	Generally, repulsive Hubbard interactions suppress the charge susceptibility, but enhance the spin susceptibility \cite{RPA1,RPA2,RPA3,Kuroki2008,Scalapino2009,Scalapino2011,Liu2013,Liu2018,Kohn:65,Raghu:10,Cho:13,Scalapino2012}. There is a critical interaction strength $U_c$, where the spin susceptibility diverges, implying the formation of spin density wave (SDW). At $U<U_c$, Cooper pairing may develop through exchanging spin and/or charge fluctuations. In particular, we consider Cooper pair scatterings both within and between the bands, hence both intra- and inter-band effective interactions $V^{\alpha\beta}(\mathbf{k,k'})$ \cite{Wu:15} (here $\alpha/\beta=1,\cdots,6$ are band indices) are accounted for.  From the effective interaction vertex $V^{\alpha\beta}(\mathbf{k,k'})$, we obtain the following linearized gap equation near the superconducting $T_c$:
	\begin{align}\label{gapeq}
		-\frac{1}{(2\pi)^3}\sum_{\beta}\oint_{FS}
		d^{2}\bm{k}'_{\Vert}\frac{V^{\alpha\beta}(\bm{k},\bm{k}')}
		{v^{\beta}_{F}(\bm{k}')}\Delta_{\beta}(\bm{k}')=\lambda
		\Delta_{\alpha}(\bm{k}).
	\end{align}
	Here the integration runs along the $\beta$- FS, $v^{\beta}_F(\bm{k}')$ is the corresponding Fermi velocity, and $\bm{k}'_\parallel$ is the component of $\bm{k}'$ along the FS. Superconducting pairing in various channels emerge as the eigenstates of the above gap equation. The leading pairing $\Delta_\alpha(\bm{k})$ is given by the eigenstate corresponding to the largest eigenvalue $\lambda$. The critical temperature $T_c$ is related to $\lambda$ through $T_c\propto e^{-1/\lambda}$.
	
	The eigenvector(s) $\Delta_{\alpha}(\bm{k})$ for each eigenvalue $\lambda$ obtained from gap equation (\ref{gapeq}) as the basis function(s) forms an irreducible representation of the $C_{6h}$ point group. In the absence of SOC, ten possible pairing symmetries are possible candidates for the system, which include five singlet pairings and five triplet pairings, as listed in Table.\ref{Tab:one}.
	
	\begin{table}
		\centering
		\caption{The ten possible pairing symmetries for K$_2$Cr$_3$As$_3$H$_x$ in the absence of SOC, among which five are spin-singlet while the rest are spin-triplet.}
		\label{Tab:one}
		\begin{tabular}{@{}ccccccccccc@{}}
			\\\hline\hline
			singlet   &&&&&&&&&&  triplet   \\
			\hline\hline
			$s$          &&&&&&&&&&  $p_z$  \\
			$(d_{x^2-y^2},d_{xy})$             &&&&&&&&&&      $(d_{x^2-y^2},d_{xy})\cdot p_z$             \\
			$(p_x,p_y)\cdot p_z$         &&&&&&&&&&  $(p_x,p_y)$          \\
			$f_{x^3-3xy^2}\cdot p_z$         &&&&&&&&&&  $f_{x^3-3xy^2}$             \\
			$f'_{y^3-3x^2y}\cdot p_z$         &&&&&&&&&&  $f'_{y^3-3x^2y}$             \\
			\hline\hline
		\end{tabular}
	\end{table}

	\section{Susceptibility}
	\label{sec:FTA}
	
	\begin{figure}[htbp]
		\centering
		\includegraphics[width=0.5\textwidth]{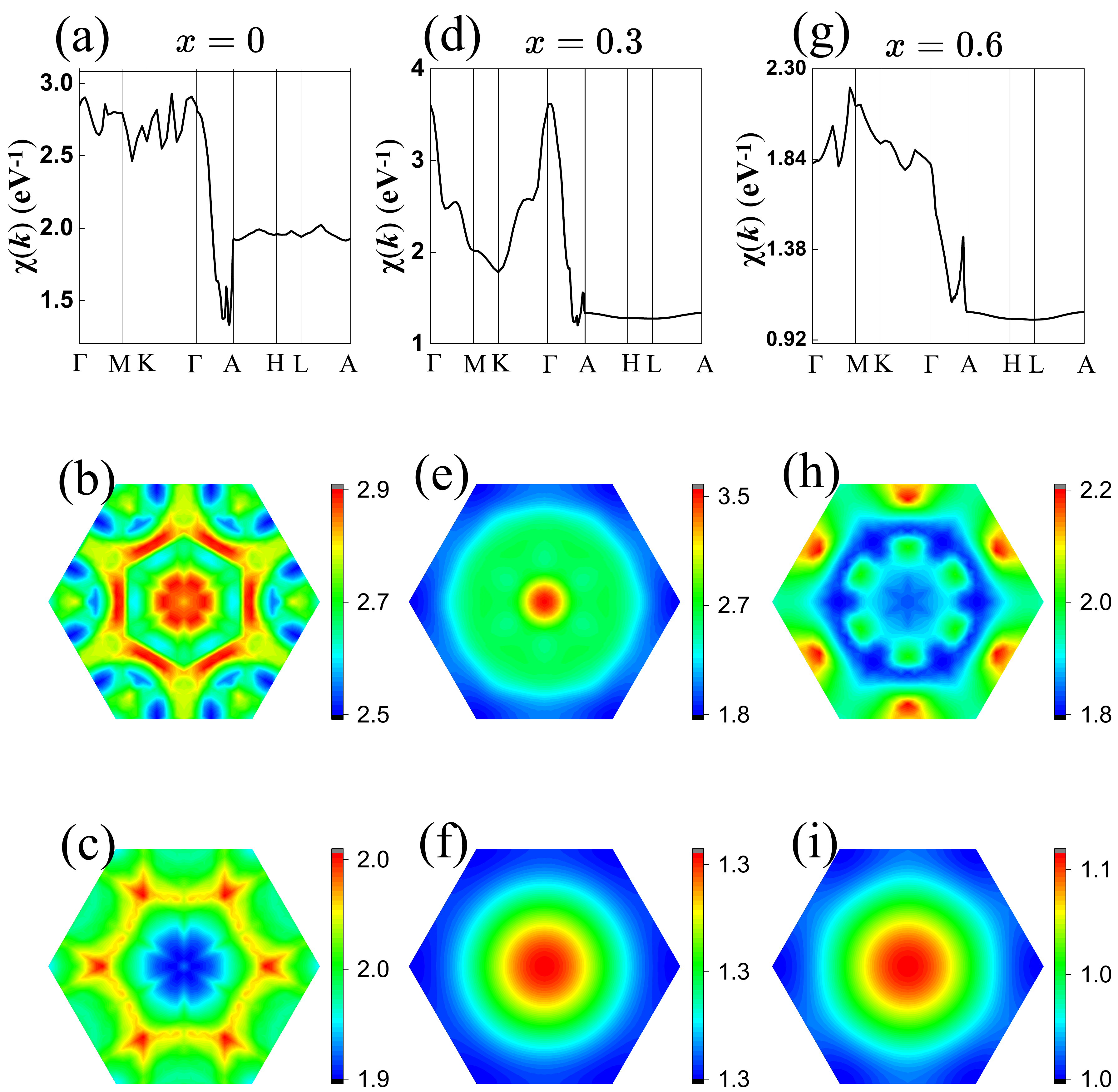}
		\caption{(color online). The $\bm{k}$-space distribution of the largest eigenvalue $\chi(\bm{k})$ of the susceptibility matrix $\chi^{(0)pq}_{st}(\bm{k},i\omega_n=0)$ for (a)-(c) K$_2$Cr$_3$As$_3$, (d)-(f) K$_2$Cr$_3$As$_3$H$_{0.3}$ and (g)-(i)  K$_2$Cr$_3$As$_3$H$_{0.6}$. From top to bottom are largest eigenvalue of the susceptibility matrix $\chi^{(0)}_{pqst}(\bm{k},i\omega_n=0)$ along the high-symmetry lines in the Brillouin zone, on the $k_z=0$ plane and on the $k_z=\pi$ plane, respectively.}
		\label{chi03}
	\end{figure}

	\begin{figure}[htbp]
		\centering
		\includegraphics[width=0.5\textwidth]{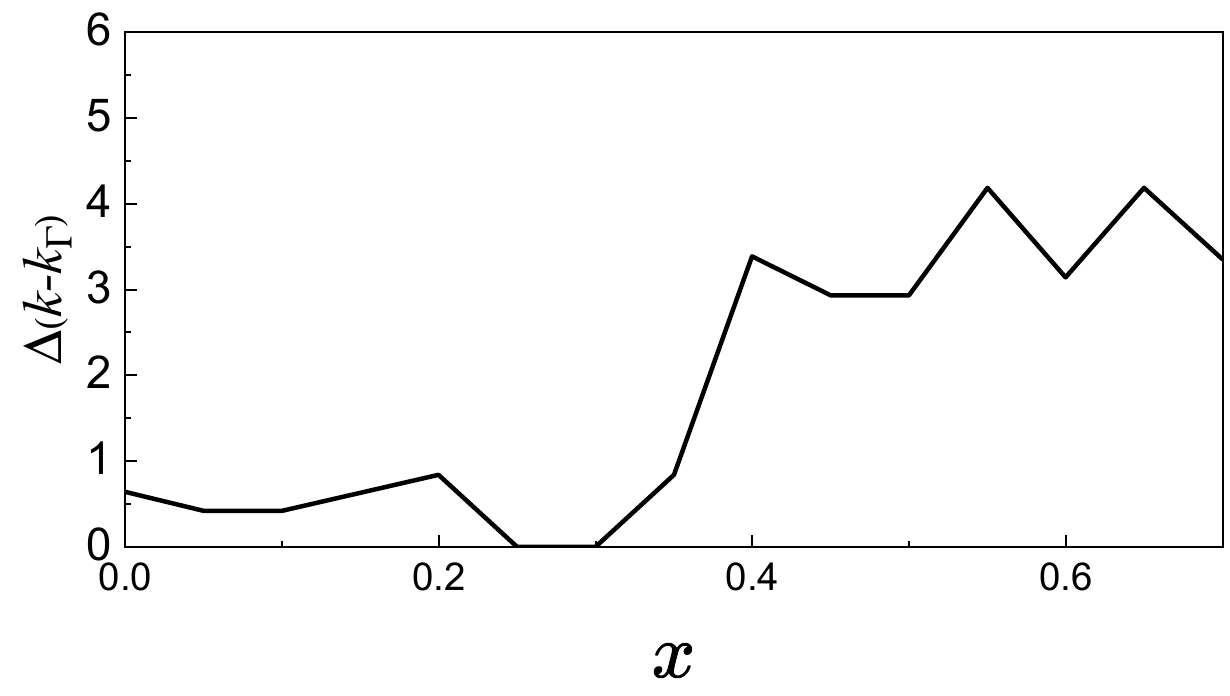}
		\caption{(color online). The hydrogen-doping level $x$ dependence of the distance between the $k$ and the $\Gamma$, where the maximum eigenvalue $\chi(\bm{k})$ of the susceptibility $\chi^{(0)pq}_{st}(\bm{k},i\omega_n=0)$ lies in.}
		\label{chig}
	\end{figure}

	The susceptibility tensor $\chi^{(0)}_{pqst}(\bm{k},i\omega)$ defined on the above can be viewed as a matrix $\chi^{(0)pq}_{st}(\bm{k},i\omega)$ by taking the combined $pq$ indices as the row index and the combined $st$ indices as the column index. In Fig.\ref{chi03}, we show the $\bm{k}$-dependence of the largest eigenvalue $\chi(\bm{k})$ of the zero-frequency susceptibility matrix $\chi^{(0)pq}_{st}(\bm{k},i\omega_n=0)$ for three different doping levels, i.e. $x=0$ in (a)-(c), $x=0.3$ in (d)-(f) and $x=0.6$ in (g)-(i). Among these figures, the (a), (d) and (g) in the first row are along the high-symmetry lines in the brillouin zone (BZ); the (b), (e) and (h) in the second row are on the $k_z=0$ plane;  and the (c), (f) and (i) in the third row are on the $k_z=\pi$ plane. Note that here $x=0$ denotes K$_2$Cr$_3$As$_3$, $x=x_c=0.3$ is the 3D-quasi-1D Lifshitz transition doping in our TB model, and the doping level $x=0.6$ is close to the highest electron-doping level we considered.

	Figure \ref{chi03} illustrates two doping-dependent features for the distributions of the susceptibilities in the BZ. The first feature lies in that the location of the $\bm{k}_M$ (which carries the strongest $\chi(\bm{k})$)  are kept in the $k_z=0$ plane from $x=0$ to $x=0.6$. We also calculated the  largest eigenvalue $\chi(\bm{k})$ for other doping points and found that in the whole doping considered, the largest eigenvalue $\chi(\bm{k})$ of the $\chi^{(0)pq}_{st}(\bm{k},i\omega_n=0)$ always fall on the $k_z=0$ plane, implying an effective FM coupling between the equivalent Cr sites along the face-sharing Cr$_6$ octahedron chains. The second feature lies in that when the doping is less than 0.3, the maximum eigenvalue $\chi(\bm{k})$ located at $\Gamma$ point, and when doping beyond 0.3, the $k$ point of the maximum eigenvalue is no longer the $\Gamma$ point. This is illustrated more clearly in Figure \ref{chig}, from which we can see the $k$ of the largest eigenvalue $\chi(\bm{k})$ is very close to $\Gamma$ point when $x\leq 0.3$, and far from $\Gamma$ when $x>0.3$. That reflects the variation from interchain ferromagnetic correlations for K$_2$Cr$_3$As$_3$ to interchain antiferromagnetic correlations for K$_2$Cr$_3$As$_3$H. Note that, although there is an FM-AFM fluctuation transition in the interchain, there is always a ferromagnetic fluctuation in the intrachain, which is consistent with previous DFT results on K$_2$Cr$_3$As$_3$H \cite{Li2023}.

	\section{Pairing phase diagram}	
	\label{phase_diagram}
	\begin{figure}[htbp]
		\centering
		\includegraphics[width=0.45\textwidth]{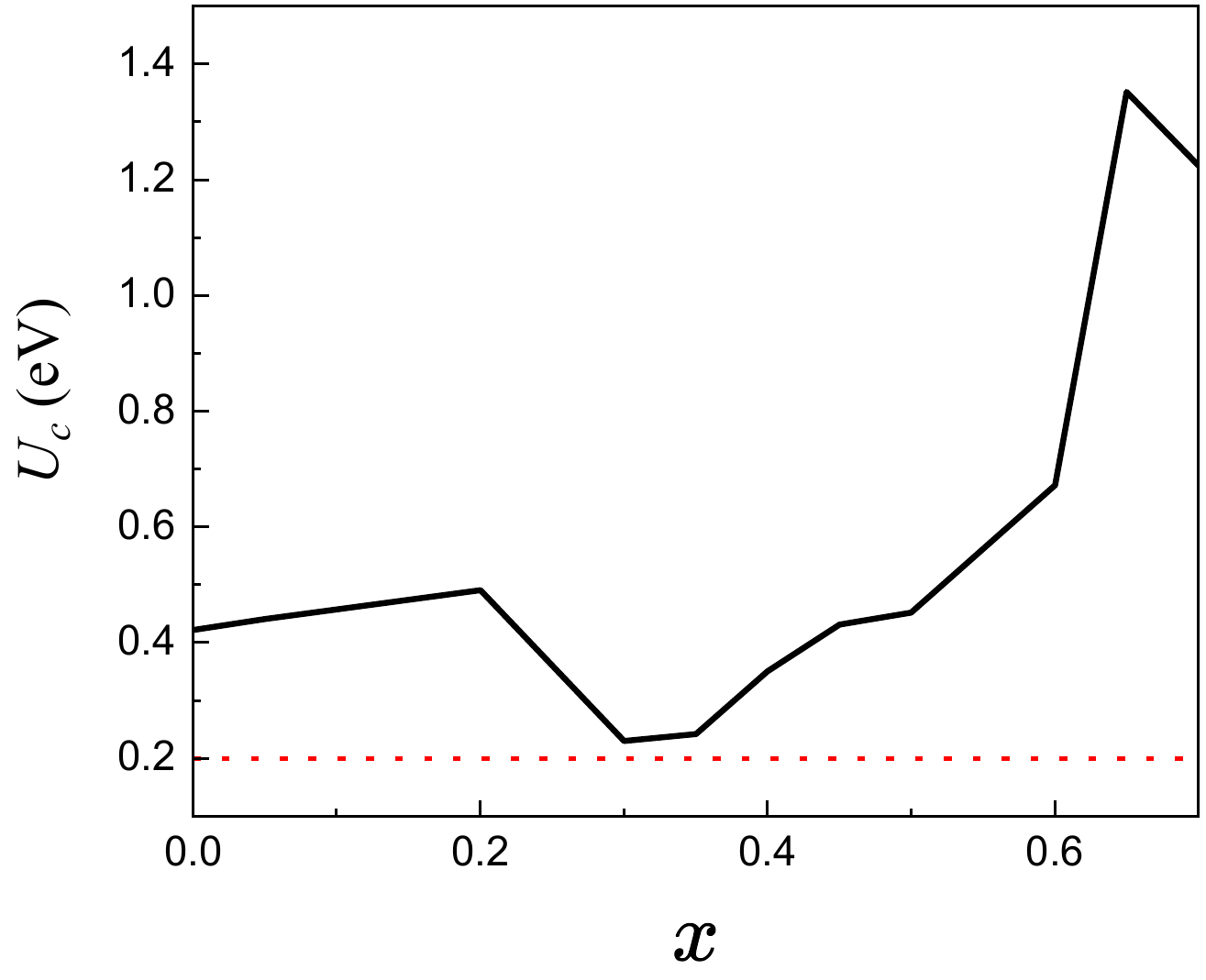}
		\caption{(color online). Critical interaction $U_c$ of the RPA spin susceptibility as a function of hole doping. In the main text, we adopt $U$=0.2 eV to calculate the pairing strength, in order to avoid magnetic instability.} 		\label{usc}
	\end{figure}

	\begin{figure}[htbp]
		\centering
		\includegraphics[width=0.45\textwidth]{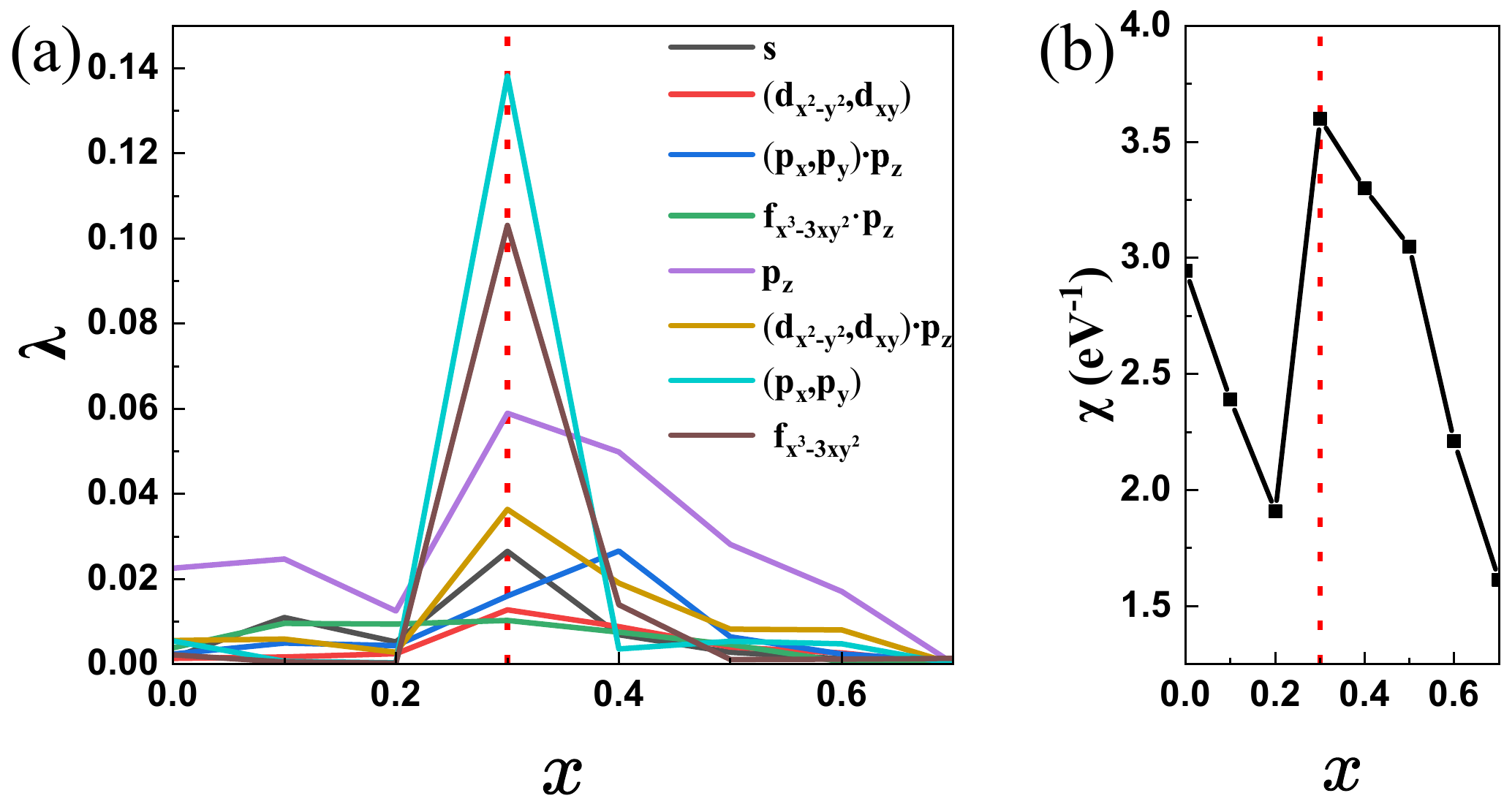}
		\caption{(color online). (a) The largest pairing eigenvalues $\lambda$ as function of $x$ for eight pairing symmetries with relatively higher $\lambda$ under $U$ = 0.2 eV, $J_H$ = 0.2$U$. (b) The doping dependences of $\chi_{Max}$.} 		\label{lambda}
	\end{figure}

	The doping dependences of the largest pairing eigenvalues $\lambda$ for various pairing symmetries are shown in Fig. \ref{lambda}(a). Eight out of the ten possible pairing symmetries with relatively higher pairing eigenvalues are shown. The parameter settings are $J_H=0.2U$ and $U=0.2$ eV, satisfying $U<U_c$ in Fig. \ref{usc}.

	Three important results are provided by Fig. \ref{lambda}. Firstly, the triplet pairing is the leading pairing symmetry in the whole doping regime of $x\in (0,0.7)$. When the doping $x$ is close to 0, the leading pairing symmetry is triplet $p_z$-wave and it transforms to be $(p_x,p_y)$-wave when $x\in (0.2,0.4)$. After $x$ exceeds 0.4, the leading pairing symmetry transforms back to be $p_z$-wave. This result, combined with previous theories \cite{Wu:15,Zhang:16,Zhou:17} and experiments \cite{Jie2021}, suggests that the SC in K$_2$Cr$_3$As$_3$H$_x$ is a robust triplet SC. From VHS in FS of these doping as shown in Fig. \ref{fs} we can find the VH momenta are located at time-reversal variant points which are called type-II VHS. It's pointed out \cite{VHS1,VHS2,VHS3,Chen2015,VHS5} that triplet SC would generally be favored near the type-II VHS. Secondly, the $T_c$ peaks near the Lifshitz-transition point with $x\approx 0.3$, which is a DOS-peak according to Fig. \ref{band}. What's more, a comparison between Fig. \ref{lambda}(a) and (b) reveals the similarity between the $\lambda\sim x$ relation for the triplet SC and the $\chi\sim x$ relation. Combine Fig. \ref{chig} we can find that the $k$ of the $\chi_{Max}$ in 0.3 doping is closest to the $\Gamma$ point. As we know, the polarization vector of the magnetic order parameters close to $\Gamma$ is favorable for ferromagnetic fluctuation. So the physical reason for the similarity between pairing symmetry and susceptibility lies in that the ferromagnetic fluctuation reflected by $\chi_{\Gamma}$ favors the formation of triplet SC. Thirdly, $(p_x,p_y)$-wave pairing occurs in this system due to hydrogen doping. This is a novel pairing symmetry of this system that we need to further analyze and understand.
	
	\section{The $p_x\pm ip_y$-wave superconductivity}	
	\label{p+ip}
	\begin{figure}[htbp]
		\centering
		\includegraphics[width=0.45\textwidth]{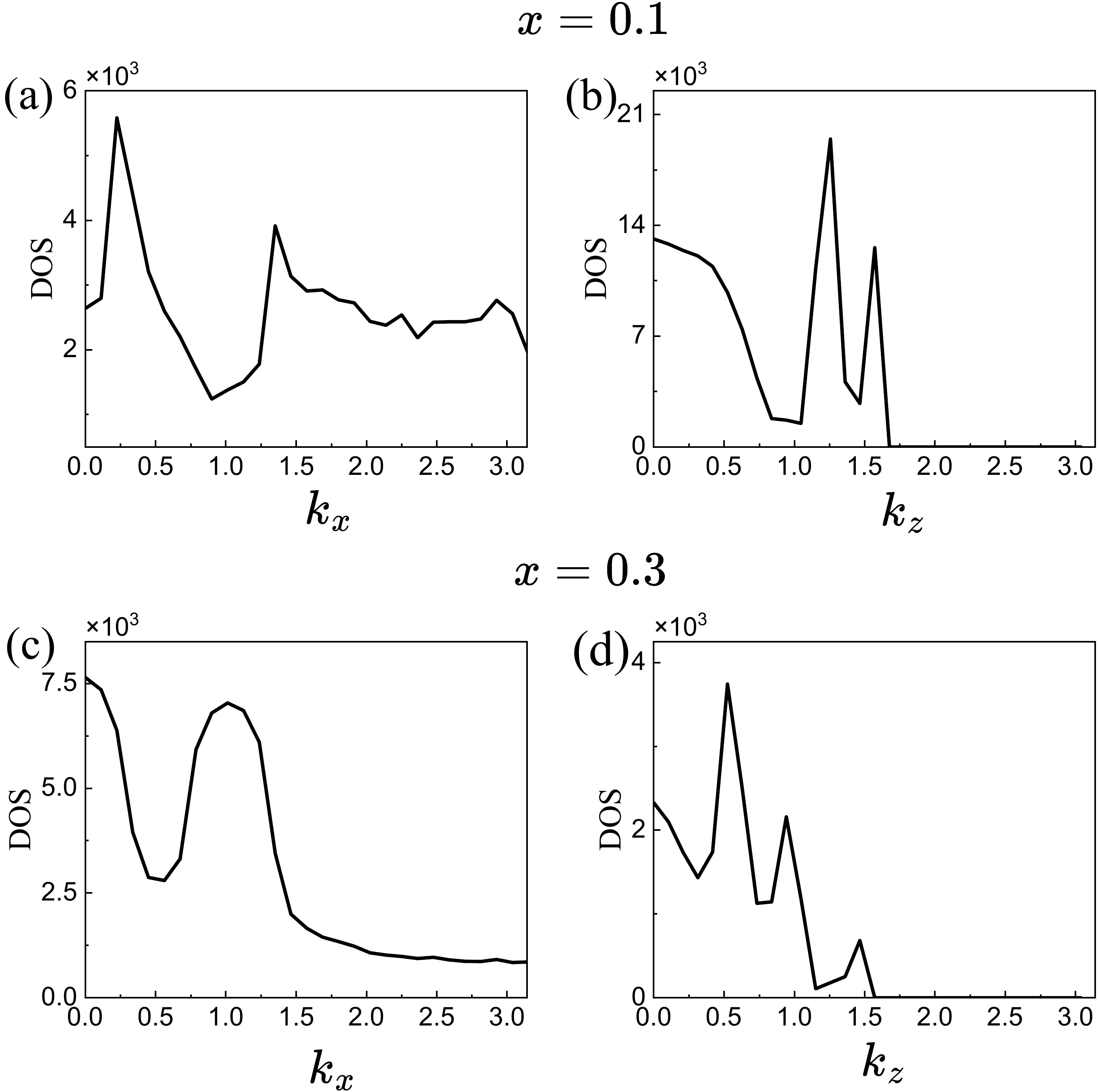}
		\caption{The $k$ dependence of DOS, where the doping is 0.1 in (a) and (b), and 0.3 in (c) and (d).} 		\label{dosk}
	\end{figure}

	The leading pairing symmetry in Fig. \ref{lambda} is $p_z$-wave in most doping levels we considered, and $(p_x,p_y)$-wave leading only in a very narrow doping level while its $T_c$ is the largest in the entire doping level. Therefore, it is necessary to check this pairing phase diagram in the physical origin. Let's start with a more thorough investigation on the detailed DOS. Fix at different doping, we consider the distribution of DOS in three orthogonal axes of the first brillouin zone. It shows significantly difference between 0.3 and other doping levels: Around 0.3 doping, DOS peaks at relatively large-$k_x(k_y)$ regime $k_x(k_y)\in (1.0,1.2)$ and relatively small-$k_z$ regime $k_z\approx 0.5$. But other doping cases are just opposite ---the DOS-peak's $k_x(k_y)$ is relatively small and $k_z$ is relatively large. In Fig. \ref{dosk}, we illustrate that for 0.1 doping case and 0.3 doping case. The reason why we focus on DOS-peak is that the regimes with relatively large DOS on the FS should be distributed with relatively large pairing gap amplitudes, so that the system can gain more energy from the superconducting condensation \cite{HuJ}. Besides, the distribution of $p$-waves is typically positively correlated with the value of $k$, e.g., $p_z$-wave pairing gap should be low in small-$k_z$ regime and be high in large-$k_z$ regime. Therefore, the pairing-wave's symmetry depends on which direction the $k$ of DOS-peak is larger. For 0.3 doping in this system, $(p_x,p_y)$-wave pairing is the leading pairing symmetry since the DOS-peak's $k_x(k_y)$ is relatively large in comparison with $k_z$ as shown in Fig. \ref{dosk}(c) and (d). For other dopings, i.e, 0.1 doping, $p_z$-wave pairing is the leading pairing symmetry since the DOS-peak's $k_z$ is relatively large in comparison with $k_x(k_y)$ as shown in Fig. \ref{dosk}(a) and (b). Thus, the pairing symmetry near 0.3 doping is $(p_x,p_y)$-wave, which is different from other doping, and the $T_c$ maximum of $(p_x,p_y)$-wave is determined by the total DOS-peak as shown in Fig. \ref{band}(b) and 3D-quasi-1D lifshitz transition take place in this doping as shown in Fig. \ref{fs}(b).
	
	\begin{figure}[htbp]
		\centering
		\includegraphics[width=0.4\textwidth]{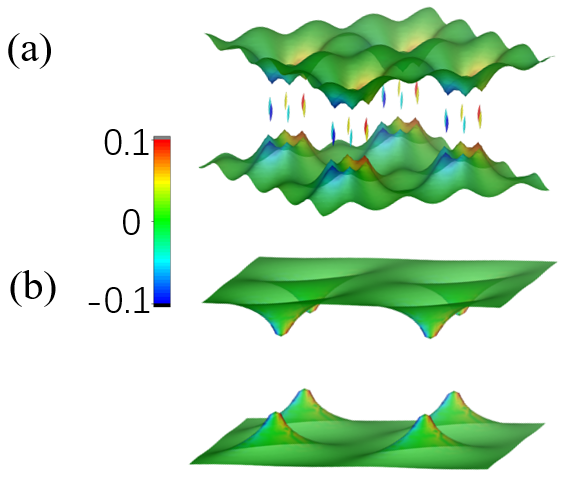}
		\caption{(color online). The pairing gap functions shown on $\alpha$ FS (a) and $\beta$ FS (b) for the $p_x\pm ip_y$-SC with $x=0.3$.} 		\label{gap}
	\end{figure}

	The distribution of the relative gap function of the obtained $p_x\pm ip_y$-wave SC is shown on the $\alpha$-, $\beta$- FSs for the Lifshitz-transition doping level $x=0.3$ in Fig.~\ref{gap} (a) and (b). While the $\alpha$- FS at this doping is quasi-1D like planes almost parallel to the $(k_x,k_y)$-plane, the $\beta$- FS has six extremely thin tubular FS sheets in the middle of the quasi-1D like planes, which arise near $k_z=0$. Figure~\ref{gap} (a) and (b) show that this gap function is mirror symmetric about the $\sigma_h$, and changes sign when rotated 180 degrees about the $z$-axis, consistent with the $p_x$ or $p_y$-wave pairing symmetry. The degenerate $p_x$- and $p_y$-wave pairings would always be mixed into the $p_x\pm ip_y$ form to lower the ground-state energy, as verified by some numerical results \cite{Lu2018} and theoretical analysis \cite{Chen2015,Liu2023}.

	From Fig.~\ref{gap} we can identify some slightly bright green gap zero points on the quasi-one-dimensional plane of the $\alpha$ and $\beta$ Fermi surfaces, which also known as nodal-point. The triplet $p_x\pm ip_y$-wave pairing state in 3D materials is so-called point-node superconductivity, as it forms points node with $k_x=k_y=0$ on the Fermi plane. While the $p_z$-wave forms line node in the $k_z=0$ plane, leading to line-node superconductivity. The experimental observable properties of point-node gap superconductivity and line-node gap superconductivity are the key criterion to distinguish $p_x\pm ip_y$-wave and $p_z$-wave. Therefore, it is necessary to investigate K$_2$Cr$_3$As$_3$H$_x$'s temperature-dependent behavior of the various experimental observable properties, e.g., the specific heat $C_v(T)$, under different $x$. Here we perform a qualitative analysis of $p_x\pm ip_y$-wave case, which will be discussed in detail in future studies.
	
	Generally, $C_v(T)$ is:
	\begin{align}\label{CV}
		C_v
		=&\frac{\partial }{\partial T}E= \frac{\partial }{\partial T}\sum_{k}\varepsilon_k^{\alpha} n_F(\varepsilon_k^{\alpha})   \nonumber\\
		=&\frac{\partial }{\partial T}\int \rho(\varepsilon)\varepsilon\frac{1}{e^{\beta\varepsilon}+1}\mathrm{d}\varepsilon \nonumber\\
		=&\int \rho(\varepsilon)\varepsilon\frac{\varepsilon e^{\beta\varepsilon}T^{-2}}{(e^{\beta\varepsilon}+1)^2}\mathrm{d}\varepsilon
	\end{align}
	Here we take the boltzmann constant $\mathrm{k}_\mathrm{B}$ as 1.
	
	When $T\to 0$, $\beta\to \infty$, $\frac{e^{\beta \varepsilon}}{(e^{\beta\varepsilon}+1)^2}$ is significantly nonzero only within the range of $\varepsilon \in (-\mathrm{k}_\mathrm{B}T,\mathrm{k}_\mathrm{B}T)$, leading to $\rho(\varepsilon)\propto \varepsilon^2$. This behavior is attributed to the presence of point nodes at $k_x=k_y=0$ in the system. Therefore, as $T\to 0$, the specific heat is:
	\begin{align}\label{CV1}
		C_v\propto& \frac{1}{T^2}\int\varepsilon^4\frac{e^{\beta\varepsilon}}{(e^{\beta\varepsilon}+1)^2}\mathrm{d}\varepsilon\nonumber\\
		=&\frac{1}{T^2}\int(\beta\varepsilon)^4\frac{e^{\beta\varepsilon}}{(e^{\beta\varepsilon}+1)^2}\mathrm{d}\beta\varepsilon\times\frac{1}{\beta^5} \nonumber\\
		\propto& T^3\int_{-\infty}^{\infty}x^4\frac{e^x}{(e^x+1)^2}\mathrm{d}x \nonumber\\
		\propto& T^3
	\end{align}
	
	Similarly, the temperature-dependent behavior of other experimentally observable properties is as follows: the Knight shift $K_{ss}\propto T^2$, the NMR spin-relaxation rate $\frac{1}{T_1T}\propto T^4$ and the superfluid density $\rho\propto T^2$ at sufficiently low temperature $T\ll T_c$, meanwhile, $C_v\propto T^2$, $K_{ss}\propto T$, $\frac{1}{T_1T}\propto T^2$ and $\rho\propto T$ in $p_z$-wave supercondutor. One can clearly judge which kind of superconducting symmetry in K$_2$Cr$_3$As$_3$H$_x$ by measuring the temperature-dependent behavior of the experimental observable properties listed above.
	
	\section{Discussion AND CONCLUSION}
	\label{sec:conclusion}

	In conclusion, adopting six-band TB model which is specified in Ref. \cite{Wu:15} equipped with the extended Hubbard interactions, we use the RPA approach to study the pairing state of the hydrogen doped K$_2$Cr$_3$As$_3$ under the rigid-band approximation. In the hydrogen-doping regime $x\in(0,0.7)$, our RPA results yield the spin triplet pairing as the leading pairing symmetry. Specifically, when $x\in(0.3,0.35)$, the leading pairing symmetry is the $p_x\pm ip_y$-wave. This is the first theoretical calculations that achieves $p_x\pm ip_y$-wave superconductivity in this system and hydrogen doping plays a key role. Since hydrogen-doped K$_2$Cr$_3$As$_3$ has already been successfully synthesized in experiments, we believe that by tuning the doping concentration, one can detect the $p_x\pm ip_y$-wave superconductivity experimentally. Interestingly, recent NMR measurements of the Knight shift show spin triplet superconductivity behavior \cite{Jie2021}. And the spin-lattice relaxation rate, $\frac{1}{T_1T}\propto T^4$ in the superconducting state, indicates the existence of point nodes in the superconducting gap function \cite{Yang:15,Luo:19}. These experimental properties indicate $p_x\pm ip_y$-wave pairing symmetry in the system \cite{Jie2021}. This may be due to the intercalation of a certain concentration of hydrogen in the experiment, as our hydrogen-doped results can support $p_x\pm ip_y$-wave superconductivity.

	Our pairing phase diagram shows that, when there is no hydrogen doping, i.e K$_2$Cr$_3$As$_3$, the triplet $p_z$-wave pairing is the leading pairing symmetry. Considering hydrogen as electron doping $x\in(0,0.7)$, $p_x\pm ip_y$-wave would be the leading pairing symmetry in $x\in (0.3,0.35)$ doping. Noteworthy, in the doping levels that the $p_x\pm ip_y$-wave leads the pairing symmetry, the $T_c$ is the highest throughout the whole doping levels being considered. Careful investigation of the physical origin we found that the $T_c$ peak is due to DOS peaking at the same doping. And the reason why the cooper pairs condensed a $p_x\pm ip_y$-wave pairing is because the DOS at this doping is dense in the larger $k_x(k_y)$. However, in the doping region of $p_z$-wave pairing, we found DOS is dense in the larger $k_z$.
	
	As for the magnetic properties, we found interchain magnetic fluctuation transforms from ferromagnetic fluctuation to antiferromagnetic fluctuation as $x$ increases. Moreover, the maximum eigenvalue of suscetibility $\chi(\bm{k})$ always falls in the $k_z=0$ plane throughout the whole doping level, which means that in-plane magnetism determines the overall magnetic performance of the system. Since it transforms into an antiferromagnetic spin fluctuation after the doping level becomes high, we can reasonably conclude that the magnetic order of K$_2$Cr$_3$As$_3$H is antiferromagnetic, which is consistent with the recent experiment \cite{Li2023}.

	Note that, although the in-plane spin susceptibility transforms from ferromagnetic fluctuation to antiferromagnetic fluctuation, the intrachain spin susceptibility is always ferromagnetic fluctuation in K$_2$Cr$_3$As$_3$H$_x$. As we know, ferromagnetic spin fluctuations favor spin-triplet Cooper pairing. Combined with our findings that show triplet pairing is leading across all doping levels, we can infer that the enhancement of spin triplet superconductivity in this quasi-1D material is due to intrachain ferromagnetic fluctuations. Further, we suppose that intra-chain magnetic fluctuations play a key role in the superconducting pairing of all quasi-one-dimensional superconductors.

	\section*{Acknowledgements}
	We are grateful to the stimulating discussions with Xianxin Wu. This work is supported by the Science Foundation of Zhejiang Sci-Tech University (ZSTU) (Grant No. 19062463-Y and No. 22062344-Y). F.Y. is supported by the National Natural Science Foundation of China under the Grants No. 12074031, No. 12234016.

\end{document}